\documentclass[dvips,apjl]{emulateapj}

\DeclareGraphicsExtensions{.eps}
\usepackage{amsmath}
\usepackage{apjfonts}

\begin{document}

\title{Exact versus approximate solutions in Gamma-Ray Burst afterglows}

\author{Carlo Luciano Bianco\altaffilmark{1,2}, Remo Ruffini\altaffilmark{1,2}}

\altaffiltext{1}{ICRA - International Center for Relativistic Astrophysics.}
\altaffiltext{2}{Dipartimento di Fisica, Università di Roma ``La Sapienza'', Piazzale Aldo Moro 5, I-00185 Roma, Italy. E-mail: bianco@icra.it, ruffini@icra.it}

\begin{abstract}
We have recently obtained the exact analytic solutions of the relativistic equations relating the radial and time coordinate of a relativistic thin uniform shell expanding in the interstellar medium in the fully radiative and fully adiabatic regimes. We here re-examine the validity of the constant-index power-law relations between the Lorentz gamma factor and its radial coordinate, usually adopted in the current Gamma-Ray Burst (GRB) literature on the grounds of an ``ultrarelativistic'' approximation. Such expressions are found to be mathematically correct but only approximately valid in a very limited range of the physical and astrophysical parameters and in an asymptotic regime which is reached only for a very short time, if any, and are shown to be not applicable to GRBs.
\end{abstract}


\section{introduction}\label{intr}

The discovery of afterglows \citep{c97} has offered a very powerful tool for reaching an understanding of Gamma-Ray Bursts (GRBs). The consensus has been reached that such an afterglow originates from a relativistic thin shell of baryonic matter propagating in the InterStellar Medium (ISM) and that its description can be obtained from the relativistic conservation laws of energy and momentum. In \citet{eqts_apjl2} we reported the exact analytic solutions of the corresponding equations, respectively under fully radiative and fully adiabatic conditions, giving in both cases explicit relations between the laboratory time and the radial coordinate of the shell. We here compare and contrast our results with the simple constant-index power-law relation between the Lorentz gamma factor and the radial coordinate of the shell generally adopted in the current literature, obtained using a so-called ``ultrarelativistic'' approximation \citep[see e.g.][and references therein]{s97,s98,w97,rm98,gps99,pm98c,pm99,cd99,p99,gw99,vpkw00,m02}. We show that such an approximation only holds in a very limited range of the physical and astrophysical parameters and in an asymptotic regime which is reached only for a very short time, if any. We demonstrate that this constant-index power-law cannot be used for modeling GRBs. Illustrative examples are given for the source GRB 991216.

\section{The afterglow analytic solutions}

The fulfillment of the energy and momentum conservation for the equations of motion of the relativistic baryonic matter shell in the laboratory reference frame leads to the following equations \citep[see e.g.][and references therein]{p99,Brasile}:
\begin{subequations}\label{Taub_Eq}
\begin{equation}
dE_{\mathrm{int}} = \left(\gamma - 1\right) dM_{\mathrm{ism}} c^2\, ,
\label{Eint}
\end{equation}
\begin{equation}
d\gamma = - \left[\left(\gamma^2 - 1\right)/M\right] dM_{\mathrm{ism}}\, ,
\label{gammadecel}
\end{equation}
\begin{equation}
dM = \left[\left(1-\varepsilon\right)/c^2\right]dE_{\mathrm{int}}+dM_\mathrm{ism}\, ,
\label{dm}
\end{equation}
\begin{equation}
dM_\mathrm{ism} = 4\pi m_p n_\mathrm{ism} r^2 dr \, ,
\label{dmism}
\end{equation}
\end{subequations}
where $M$ is the shell mass-energy, $n_\mathrm{ism}$ is the ISM number density, $m_p$ is the proton mass, $\varepsilon$ is the emitted fraction of the energy developed in the collision with the ISM and $M_\mathrm{ism}$ is the amount of ISM mass swept up by the shell within the radius $r$: $M_\mathrm{ism}=(4\pi/3)m_pn_\mathrm{ism}(r^3-{r_\circ}^3)$, where $r_\circ$ is the starting radius of the baryonic matter shell. In general, an additional equation is needed in order to express the dependence of $\varepsilon$ on the radial coordinate. In the following, $\varepsilon$ is assumed to be constant and such an approximation appears to be correct in the GRB context.

Consensus has also been reached on a simple integration of the equations of motion Eqs.\eqref{Taub_Eq} in the fully radiative case $(\varepsilon=1)$ \citep[see][]{p99,Brasile,eqts_apjl2}), leading to:
\begin{subequations}\label{analsol}
\begin{equation}
\gamma=\frac{1+\left(M_\mathrm{ism}/M_B\right)\left(1+\gamma_\circ^{-1}\right)\left[1+\tfrac{1}{2}\left(M_\mathrm{ism}/M_B\right)\right]}{\gamma_\circ^{-1}+\left(M_\mathrm{ism}/M_B\right)\left(1+\gamma_\circ^{-1}\right)\left[1+\tfrac{1}{2}\left(M_\mathrm{ism}/M_B\right)\right]}\, ,
\label{dg_exp}
\end{equation}
where $M_B$ and $\gamma_\circ$ are the initial values respectively of the mass and of the Lorentz gamma factor of the baryonic shell. The above Eqs.\eqref{Taub_Eq},\eqref{dg_exp} differ from the ones derived by \citet{bm76} in a different framework, often quoted in the current literature within the present context. Correspondingly, in the fully adiabatic case $(\varepsilon=0)$, Eqs.\eqref{Taub_Eq} have the following analytic solution \citep[see][]{p99,eqts_apjl2}):
\begin{equation}
\gamma^2=\frac{\gamma_\circ^2+2\gamma_\circ\left(M_\mathrm{ism}/M_B\right)+\left(M_\mathrm{ism}/M_B\right)^2}{1+2\gamma_\circ\left(M_\mathrm{ism}/M_B\right)+\left(M_\mathrm{ism}/M_B\right)^2}\, .
\label{dg_exp_ad}
\end{equation}
\end{subequations}

In \citet{eqts_apjl2} we have explicitly integrated the differential equations for $r(t)$ in Eqs.\eqref{analsol}, recalling that $\gamma^{-2}=1-[dr/(cdt)]^2$, where $t$ is the time in the laboratory reference frame. We have then obtained new explicit analytic expressions for the equations of motion of the relativistic shell which are essential for explicitly obtaining the analytic expressions of the equitemporal surfaces (EQTSs) in the fully radiative and in the fully adiabatic cases, respectively \citep[see][]{eqts_apjl,eqts_apjl2}.

\section{Approximations adopted in the current literature}

We turn now to the comparison of the exact solutions given in Eqs.\eqref{analsol} with the approximations used in the current literature. Following \citet{bm76}, a so-called ``ultrarelativistic'' approximation $\gamma_\circ \gg \gamma \gg 1$ has been widely adopted by many authors to solve Eqs.\eqref{Taub_Eq} (see references in section \ref{intr}). This leads to simple constant-index power-law relations:
\begin{equation}
\gamma\propto r^{-a}\, ,
\label{gr0}
\end{equation}
with $a=3$ in the fully radiative case and $a=3/2$ in the fully adiabatic case. This simple relation is in contrast with the complexity of Eqs.\eqref{analsol}.

We address now the issue of establishing the domain of applicability of the simplified Eq.\eqref{gr0} used in the current literature both in the fully radiative and adiabatic cases.

\subsection{The fully radiative case}\label{fr}

We first consider the fully radiative case. If we assume:
\begin{equation}
1/\left(\gamma_\circ+1\right) \ll M_\mathrm{ism}/M_B \ll \gamma_\circ/\left(\gamma_\circ+1\right) < 1\, ,
\label{app_FB}
\end{equation}
we have that in the numerator of Eq.\eqref{dg_exp} the linear term in $M_\mathrm{ism}/M_B$ is negligible with respect to $1$ and the quadratic term is {\em a fortiori} negligible, while in the denominator the linear term in $M_\mathrm{ism}/M_B$ is the leading one. Eq.\eqref{dg_exp} then becomes:
\begin{equation}
\gamma\simeq\left[\gamma_\circ/\left(\gamma_\circ+1\right)\right] M_B/M_\mathrm{ism}\, .
\label{dg_exp_2}
\end{equation}
If we multiply the terms of Eq.\eqref{app_FB} by $(\gamma_\circ+1)/\gamma_\circ$, we obtain $1/\gamma_\circ\ll(M_\mathrm{ism}/M_B)[(\gamma_\circ+1)/\gamma_\circ]\ll 1$, which is equivalent to $\gamma_\circ\gg[\gamma_\circ/(\gamma_\circ+1)](M_B/M_\mathrm{ism})\gg 1$, or, using Eq.\eqref{dg_exp_2}, to:
\begin{equation}
\gamma_\circ\gg\gamma\gg 1\, ,
\label{cond_rad}
\end{equation}
which is indeed the inequality adopted in the ``ultrarelativistic'' approximation in the current literature. If we further assume $r^3 \gg r_\circ^3$, Eq.\eqref{dg_exp_2} can be further approximated by a simple constant-index power-law as in Eq.\eqref{gr0}:
\begin{equation}
\gamma \simeq \left[\gamma_\circ/\left(\gamma_\circ+1\right)\right] M_B/\left[\left(4/3\right)\pi n_\mathrm{ism}m_pr^3\right]\, \propto\, r^{-3}\, .
\label{dg_exp_3}
\end{equation}

We turn now to the range of applicability of these approximations, consistently with the inequalities given in Eq.\eqref{app_FB}. It then becomes manifest that these inequalities can only be enforced in a finite range of $M_\mathrm{ism}/M_B$. The lower limit (LL) and the upper limit (UL) of such range can be conservatively estimated:
\begin{subequations}\label{LL-UL}
\begin{equation}
\left(\tfrac{M_\mathrm{ism}}{M_B}\right)_{LL}=10^2\tfrac{1}{\gamma_\circ+1}\, ,\quad \left(\tfrac{M_\mathrm{ism}}{M_B}\right)_{UL}=10^{-2}\tfrac{\gamma_\circ}{\gamma_\circ+1}\, .
\label{LL}
\end{equation}
The allowed range of variability, if it exists, is then given by:
\begin{equation}
\left(\tfrac{M_\mathrm{ism}}{M_B}\right)_{UL}-\left(\tfrac{M_\mathrm{ism}}{M_B}\right)_{LL}=10^{-2}\tfrac{\gamma_\circ-10^4}{\gamma_\circ+1}>0\, .
\label{delta}
\end{equation}
\end{subequations}
A necessary condition for the applicability of the above approximations is therefore:
\begin{equation}
\gamma_\circ > 10^4\, .
\label{g0}
\end{equation}
It is important to emphasize that Eq.\eqref{g0} is only a {\em necessary} condition for the applicability of the approximate Eq.\eqref{dg_exp_3}, but it is not {\em sufficient}: Eq.\eqref{dg_exp_3} in fact can be applied only in the very limited range of $r$ values whose upper and lower limits are given in Eq.\eqref{LL}. See for explicit examples section \ref{exampl} below.

\subsection{The adiabatic case}\label{ad}

We now turn to the adiabatic case. If we assume:
\begin{equation}
1/\left(2\gamma_\circ\right)\ll M_\mathrm{ism}/M_B \ll \gamma_\circ/2\, ,
\label{app_FB_ad}
\end{equation}
we have that in the numerator of Eq.\eqref{dg_exp_ad} all terms are negligible with respect to $\gamma_\circ^2$, while in the denominator the leading term is the linear one in $M_\mathrm{ism}/M_B$. Eq.\eqref{dg_exp_ad} then becomes:
\begin{equation}
\gamma\simeq\sqrt{\left(\gamma_\circ/2\right) M_B/M_\mathrm{ism}}\, .
\label{dg_exp_ad_2}
\end{equation}
If we multiply the terms of Eq.\eqref{app_FB_ad} by $2/\gamma_\circ$, we obtain $1/\gamma_\circ^2\ll(2/\gamma_\circ)(M_\mathrm{ism}/M_B)\ll 1$, which is equivalent to $\gamma_\circ^2\gg(\gamma_\circ/2)(M_B/M_\mathrm{ism})\gg 1$, or, using Eq.\eqref{dg_exp_ad_2}, to:
\begin{equation}
\gamma_\circ^2\gg\gamma^2\gg 1\, .
\label{cond_ad}
\end{equation}
If we now further assume $r^3 \gg r_\circ^3$, Eq.\eqref{dg_exp_ad_2} can be further approximated by a simple constant-index power-law as in Eq.\eqref{gr0}:
\begin{equation}
\gamma \simeq \sqrt{\left(\gamma_\circ/2\right) M_B/\left[\left(4/3\right)\pi n_\mathrm{ism}m_pr^3\right]}\, \propto\, r^{-3/2}\, .
\label{dg_exp_ad_3}
\end{equation}

We turn now to the range of applicability of these approximations, consistently with the inequalities given in Eq.\eqref{app_FB_ad}. It then becomes manifest that these inequalities can only be enforced in a finite range of $M_\mathrm{ism}/M_B$. The lower limit (LL) and the upper limit (UL) of such range can be conservatively estimated:
\begin{subequations}\label{LL-UL_ad}
\begin{equation}
\left(\tfrac{M_\mathrm{ism}}{M_B}\right)_{LL}=10^2\tfrac{1}{2\gamma_\circ}\, ,\quad \left(\tfrac{M_\mathrm{ism}}{M_B}\right)_{UL}=10^{-2}\tfrac{\gamma_\circ}{2}\, .
\label{LL_ad}
\end{equation}
The allowed range of variability, if it exists, is then given by:
\begin{equation}
\left(\tfrac{M_\mathrm{ism}}{M_B}\right)_{UL}-\left(\tfrac{M_\mathrm{ism}}{M_B}\right)_{LL}=10^{-2}\tfrac{\gamma_\circ^2-10^4}{2\gamma_\circ}>0\, .
\label{delta_ad}
\end{equation}
\end{subequations}
A necessary condition for the applicability of the above approximations is therefore:
\begin{equation}
\gamma_\circ > 10^2\, .
\label{g0_ad}
\end{equation}
Again, it is important to emphasize that Eq.\eqref{g0_ad} is only a {\em necessary} condition for the applicability of the approximate Eq.\eqref{dg_exp_ad_3}, but it is not {\em sufficient}: Eq.\eqref{dg_exp_ad_3} in fact can be applied only in the very limited range of $r$ values whose upper and lower limits are given in Eq.\eqref{LL_ad}. See for explicit examples section \ref{exampl} below.

\section{A specific example}\label{exampl}

\begin{figure}
\centering
\includegraphics[width=\hsize,clip]{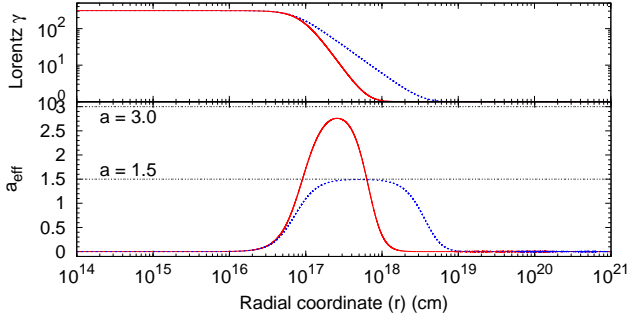}
\caption{In the upper panel, the analytic behavior of the Lorentz $\gamma$ factor during the afterglow era is plotted versus the radial coordinate of the expanding thin baryonic shell in the fully radiative case of GRB 991216 (solid red line) and in the adiabatic case starting from the same initial conditions (dotted blue line). In the lower panel are plotted the corresponding values of the ``effective'' power-law index $a_{eff}$ (see Eq.\eqref{eff_a}), which is clearly not constant, is highly varying and systematically lower than the constant values $3$ and $3/2$ purported in the current literature (horizontal dotted black lines).}
\label{2gamma}
\end{figure}

Having obtained the analytic expression of the Lorentz gamma factor for the fully radiative case in Eq.\eqref{dg_exp}, we illustrate in Fig. \ref{2gamma} the corresponding gamma factor as a function of the radial coordinate in the afterglow era for GRB 991216 \citep[see][and references therein]{Brasile}. We have also represented the corresponding solution which can be obtained in the adiabatic case, using Eq.\eqref{dg_exp_ad}, starting from the same initial conditions. It is clear that in both cases there is not a simple power-law relation like Eq.\eqref{gr0} with a constant index $a$. We can at most define an ``instantaneous'' value $a_{eff}$ for an ``effective'' power-law behavior:
\begin{equation}
a_{eff} = - \frac{d\ln\gamma}{d\ln r}\, .
\label{eff_a}
\end{equation}
Such an ``effective'' power-law index of the exact solution smoothly varies from $0$ to a maximum value which is always smaller than $3$ or $3/2$, in the fully radiative and adiabatic cases respectively, and finally decreases back to $0$ (see Fig. \ref{2gamma}). We see in particular, from Fig. \ref{2gamma}, how in the fully radiative case the power-law index is consistently smaller than $3$, and in the adiabatic case $a_{eff} = 3/2$ is approached only for a small interval of the radial coordinate corresponding to the latest parts of the afterglow with a Lorentz gamma factor of the order of $10$. In this case of GRB 991216 we have, in fact, $\gamma_\circ = 310.13$ and neither Eq.\eqref{cond_rad} nor Eq.\eqref{cond_ad} can be satisfied for any value of $r$. Therefore, neither in the fully radiative nor in the adiabatic case the constant-index power-law expression in Eq.\eqref{gr0} can be applied.

\begin{figure*}
\centering
\includegraphics[width=8.9cm,clip]{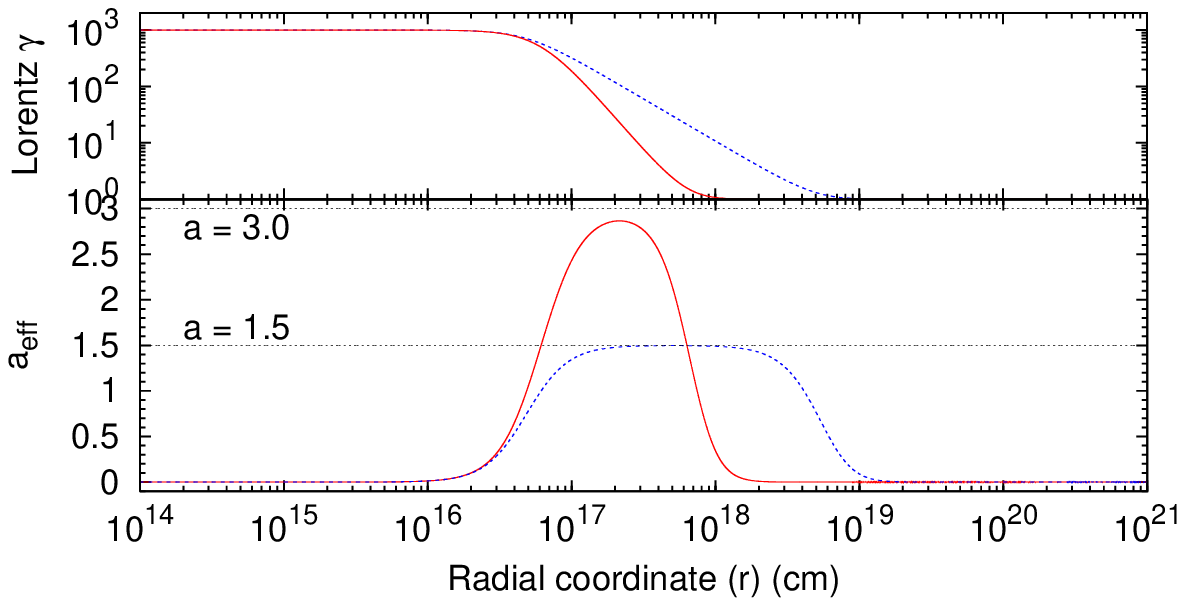}
\includegraphics[width=8.9cm,clip]{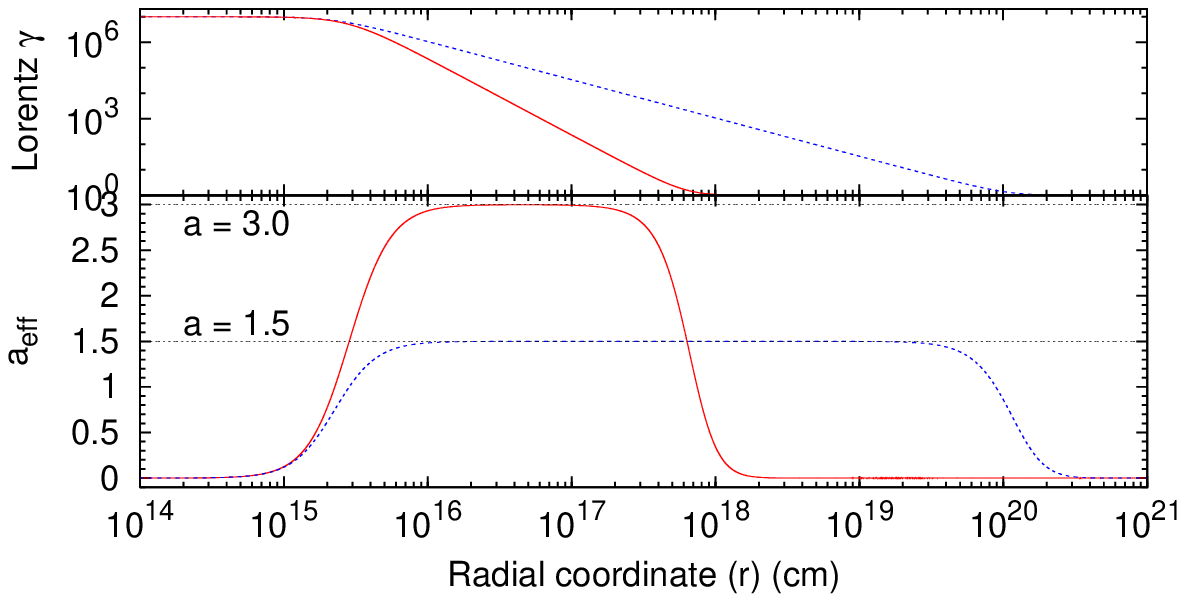}
\includegraphics[width=8.9cm,clip]{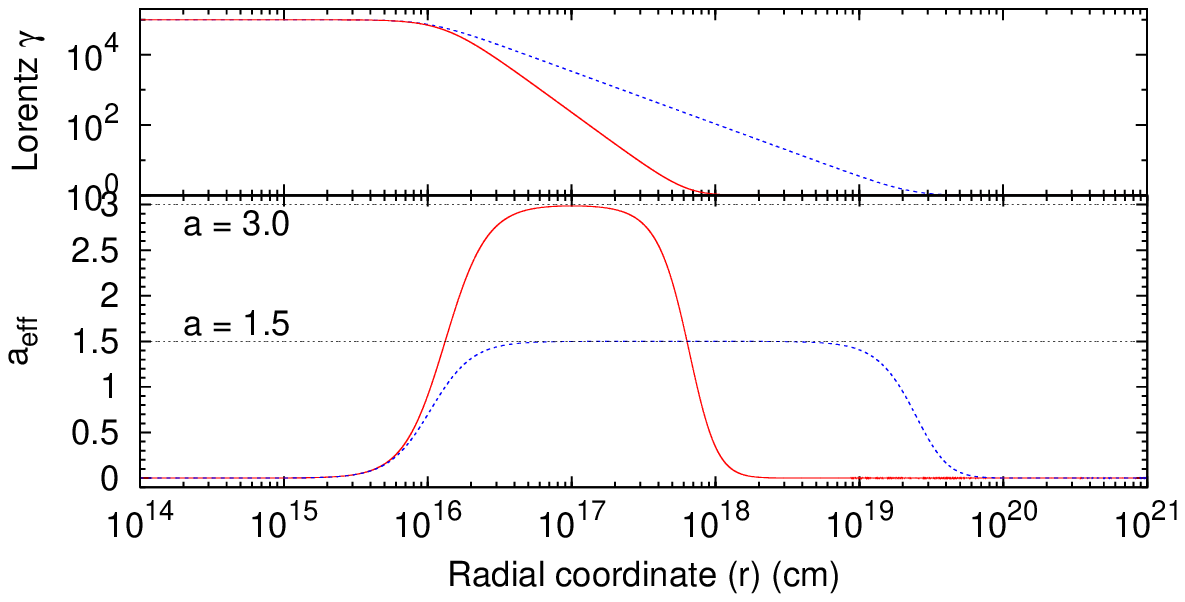}
\includegraphics[width=8.9cm,clip]{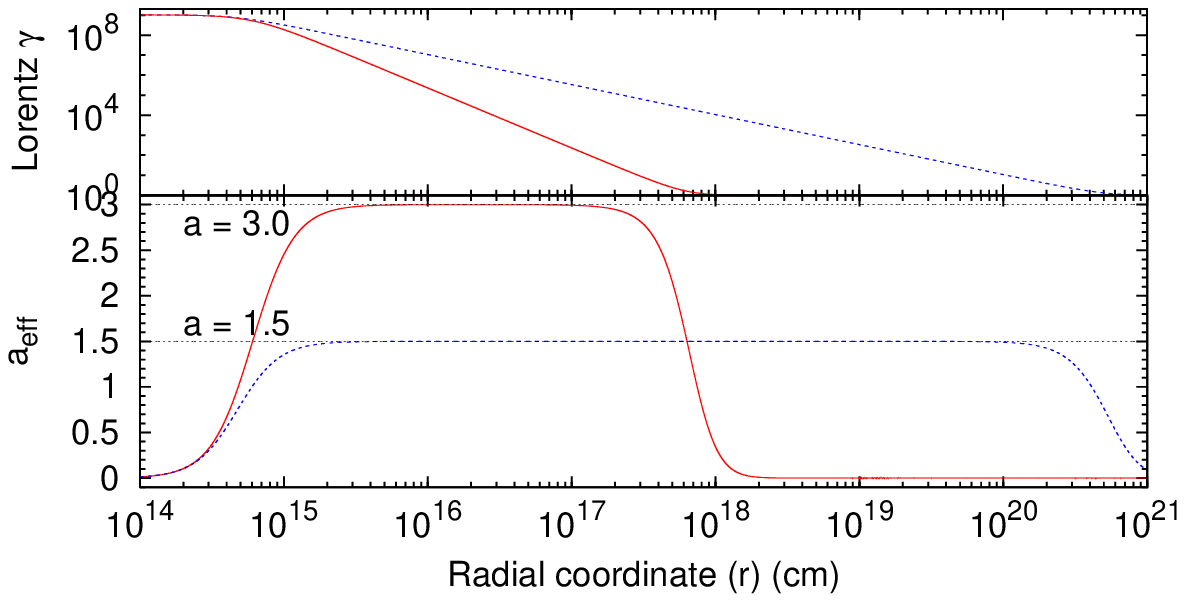}
\caption{In these four diagrams we reproduce the same quantities plotted in Fig. \ref{2gamma} for four higher values of $\gamma_\circ$. The upper (lower) left diagram corresponds to $\gamma_\circ = 10^3$ ($\gamma_\circ = 10^5$). The upper (lower) right diagram corresponds to $\gamma_\circ = 10^7$ ($\gamma_\circ = 10^9$). It is manifest how asymptotically, by increasing the value of $\gamma_\circ$, the values $a = 3$ and $a = 3/2$ (horizontal black dotted lines) are reached, but only in a limited range of the radial co-ordinate and anyway for values of $\gamma_\circ$ much larger than the ones actually observed in GRBs.}
\label{multigamma}
\end{figure*}

For clarity, we have integrated in Fig. \ref{multigamma} an ideal GRB afterglow with the initial conditions as GRB 991216 for selected higher values of the initial Lorentz gamma factor: $\gamma_\circ = 10^3, 10^5, 10^7, 10^9$. For $\gamma_\circ = 10^3$, we then see that, again, in the fully radiative condition $a_{eff} = 3$ is never reached and in the adiabatic case $a_{eff} \simeq 3/2$ only in the region where $10 < \gamma < 50$. Similarly, for $\gamma_\circ = 10^5$, in the fully radiative case $a_{eff} \simeq 3$ is only reached around the point $\gamma = 10^2$, and in the adiabatic case $a_{eff} \simeq 3/2$ for $10 < \gamma < 10^2$, although the non-power-law behavior still remains in the early and latest afterglow phases corresponding to the $\gamma \equiv \gamma_\circ$ and $\gamma \to 1$ regimes. The same conclusion can be reached for the remaining cases $\gamma_\circ = 10^7$ and $\gamma_\circ = 10^9$.

We like to emphasize that the early part of the afterglow, where $\gamma \equiv \gamma_\circ$, which cannot be described by the constant-index power-law approximation, do indeed corresponds to the rising part of the afterglow bolometric luminosity and to its peak, which is reached as soon as the Lorentz gamma factor starts to decrease. We have shown \citep[see e.g.][and references therein]{lett2,Brasile,Brasile2} how the correct identifications of the raising part and the peak of the afterglow are indeed crucial for the explanation of the observed ``prompt radiation''. Similarly, the power-law cannot be applied during the entire approach to the newtonian regime, which corresponds to some of the actual observations occurring in the latest afterglow phases.

\section{Conclusions}

It is well known that scaling laws and constant-index power-law expressions are obtainable only in the asymptotic case of ultrarelativistic regimes and in the Newtonian limit, while in the fully relativistic regime the scaling laws break down \citep[see e.g.][]{r73}. This circumstance is more subtle in GRB afterglows: 1) the ultrarelativistic approximation is only a necessary condition, but {\em not} a sufficient one, for the existence of scaling laws; 2) such a necessary condition implies values of the initial Lorentz gamma factor $\gamma_\circ$ outside the range currently observed in GRB sources.

We have shown in section \ref{fr} that, in the fully radiative case, the {\em necessary} ultrarelativistic condition for obtaining the appearance of scaling laws is $\gamma_\circ > 10^4$. We recall that the $\gamma_\circ$ values deduced typically for GRBs are of the order of $\gamma_\circ\simeq 10^2$ \citep[see e.g. GRB~030329, GRB~020322, GRB~991216, GRB~980519, GRB~980425, GRB~970228][and references therein]{cospar02,Brasile,r03mg10}. Thus this necessary condition is never fulfilled in GRBs.

It would appear from section \ref{ad} that the {\em necessary} ultrarelativistic condition for obtaining the appearance of scaling laws is less severe in the adiabatic case: $\gamma_\circ > 10^2$. However, this condition is not {\em sufficient} for the applicability of the constant-index power-law approximation to the entire afterglow, as clearly shown in Fig. \ref{2gamma}. The regime $a_{eff} = 3/2$ is in fact approached only asymptotically and anyway in a very limited region.

In the current literature (see references in section \ref{intr}), a systematic use of the constant-index power-law approximation for the Lorentz gamma factor has been made. The $\gamma \equiv \gamma_\circ$ regime has been generally neglected or erroneously matched to the constant-index power-law approximation hampering the understanding of the observed ``prompt radiation'' \citep[see e.g.][]{lett2,Brasile,Brasile2}.

We expect that the data from Swift will soon add observational evidence to the validity of this theoretical treatment.

\acknowledgments

We thank S. Blinnikov, J. Ehlers and L. Titarchuk for discussions on the wording of our manuscript.

\end{document}